\documentclass{article}

\usepackage{arxiv}

\usepackage[utf8]{inputenc} 
\usepackage[T1]{fontenc}    
\usepackage{hyperref}       
\usepackage{url}            
\usepackage{booktabs}       
\usepackage{amsfonts}       
\usepackage{nicefrac}       
\usepackage{microtype}      
\usepackage{lipsum}

\usepackage{relsize}
\usepackage{verbatim}
\usepackage{graphicx}
\graphicspath{{Figures/}}
\usepackage{dcolumn}
\usepackage{bm}
\usepackage{subcaption}
\usepackage{float}

\usepackage{amsmath,amssymb,amsfonts}
\usepackage{graphics}
\usepackage{color}
\usepackage{xcolor}
\definecolor{rev1}{rgb}{0,0,0}

\usepackage{cite}

\usepackage[ruled,vlined]{algorithm2e}
\title{Nonlinear proper orthogonal decomposition for convection-dominated flows}

\author{
 Shady E. Ahmed \\
  School of Mechanical \& Aerospace Engineering,\\
  Oklahoma State University,\\
  Stillwater, OK 74078, USA.\\
  \texttt{shady.ahmed@okstate.edu}
\And
 Omer San \\
  School of Mechanical \& Aerospace Engineering,\\
  Oklahoma State University,\\
  Stillwater, OK 74078, USA.\\
  \texttt{osan@okstate.edu} 
  \And
 Adil Rasheed\\
  Department of Engineering Cybernetics,\\
  Norwegian University of Science and Technology,\\
  N-7465, Trondheim, Norway.\\
  Department of Mathematics and Cybernetics,\\
  SINTEF Digital,\\ 7034 Trondheim, Norway. \\
\texttt{adil.rasheed@ntnu.no }
\And
 Traian Iliescu \\
  Department of Mathematics,\\
  Virginia Tech,\\
  Blacksburg, VA 24061, USA.\\
  \texttt{iliescu@vt.edu } 
}

\begin{document}
\maketitle

\begin{abstract}
Autoencoder techniques find increasingly common use in reduced order modeling as a means to create a latent space. This reduced order representation offers a modular data-driven modeling approach for nonlinear dynamical systems when integrated with a time series predictive model. In this letter, we put forth a nonlinear proper orthogonal decomposition (POD) framework, which is an end-to-end Galerkin-free model combining autoencoders with long short-term memory networks for dynamics. By eliminating the projection error due to the truncation of Galerkin models, a key enabler of the proposed nonintrusive approach is the kinematic construction of a nonlinear mapping between the full-rank expansion of the POD coefficients and the latent space where the dynamics evolve. We test our framework for model reduction of a convection-dominated system, which is generally challenging for reduced order models. Our approach not only improves the accuracy, but also significantly reduces the computational cost of training and testing. 
\end{abstract}

\keywords{Reduced order models, data-driven modeling, autoencoders, long short-term memory networks} 

\section{Introduction}
Full order models (FOMs) based on the solution of the Navier-Stokes equations, e.g.,  direct numerical simulations, large eddy simulations, and Reynolds averaged Navier-Stokes simulations, have made a tremendous impact in the numerical simulation of high Reynolds number fluid flows. However, current FOMs cannot be used effectively in multiple-query simulations (e.g., uncertainty quantification, optimization, and control) because of their prohibitively high computational cost. Reduced order models (ROMs), on the other hand, are efficient low-dimensional models created from available data. ROMs have been often used as surrogate models for structure-dominated problems. However, although traditional projection-based ROMs work in simple, canonical test problems, they generally fail (or require a significantly large number of basis functions) for convection-dominated flows because a low-dimensional ROM basis cannot accurately represent the complex dynamics.

\begin{figure*}[t!]
\centering
\includegraphics[width=0.99\textwidth]{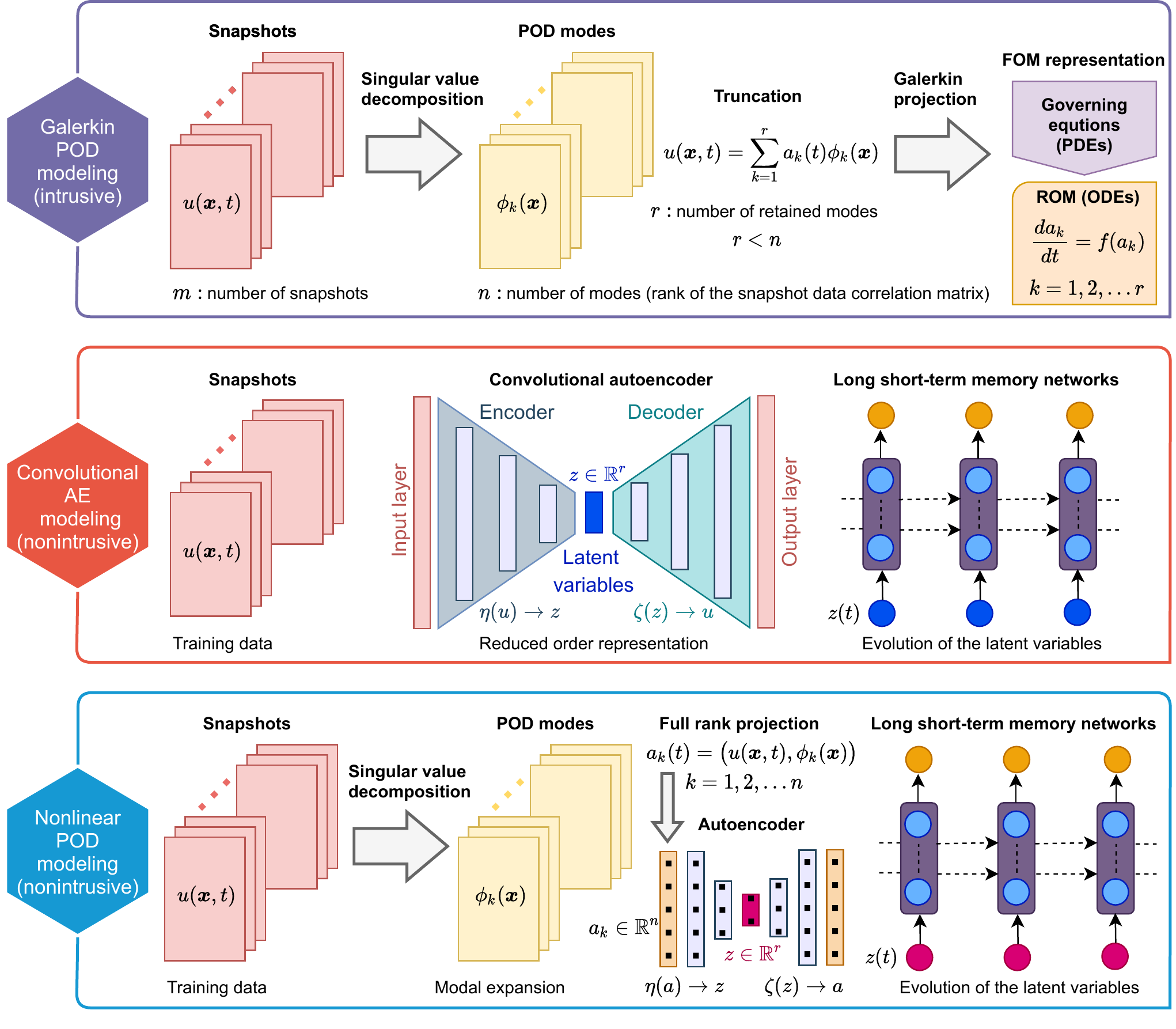}
\caption{Illustration of the research methodology: (i) Galerkin POD modeling -- an intrusive approach, which has been the workhorse for most projection-based ROM applications, (ii) convolutional autoencoder modeling -- a modular nonintrusive ROM framework that can be built from data without requiring access to the underlying governing equations, and (iii) the proposed nonlinear POD modeling approach that utilizes the autoencoder technology on a manifold defined by the full-rank POD coefficients (i.e., considering 99.9\% of the relative information content). The elimination of the subspace projection error stemming from the truncation is a major advantage of the proposed nonlinear POD method over the Galerkin POD approach, especially for the nonlinear dynamical systems with slow decay in the Kolmogorov N-width, such as convection-dominated flows studied in this paper. The proposed nonintrusive approach has also two other advantages over the convolutional autoencoder model. First, it can reduce the computational cost of the training process since it involves significantly fewer trainable parameters. Second, it ensures an accurate implementation of the boundary conditions since the spatial variations are naturally embedded in the reduced order representation. }
\label{fig:nonPOD}
\end{figure*}

Among fluid dynamicists, projection-based truncated methods are quite popular. Practitioners often utilize the proper orthogonal decomposition (POD) method to generate a set of orthonormal basis functions, or a reduced order representation \cite{ahmed2021closures}. After a suitable basis is chosen, the Galerkin projection is generally used to project the dynamics of the equation onto the subspace spanned by a truncated set of basis functions (e.g., see the Galerkin POD (GPOD) approach in the top panel of Figure~\ref{fig:nonPOD}).  Although time-periodic or quasi-time-periodic dynamical systems can be easily represented by a small number of POD modes, convection-dominated unsteady flows might require a large number of POD modes. Therefore, a significant projection error is often introduced by a truncation process. This projection error increases for dynamical systems when the Kolmogorov N-width \cite{greif2019decay,ahmed2020breaking} increases. \textcolor{rev1}{The Kolmogorov N-width is an approximation theory concept that determines the linear reducibility of the underlying systems, which can be connected to the POD spectrum \cite{djouadi2010connection}.}

The emerging autoencoder (AE) technology provides a powerful alternative way of generating a reduced order representation, often called latent space or finite-dimensional manifold. In recent years, there has been an ever-increasing interest in discovering inertial manifolds of partial differential equations (PDEs) using autoencoders \cite{champion2019data,linot2020deep,sondak2021learning,mojgani2021low,kim2020efficient}. Convolutional AE (CAE) is a promising approach that has been utilized in fluid dynamics applications \cite{gonzalez2018learning,wiewel2019latent,mohan2019compressed,bhatnagar2019prediction,murata2020nonlinear,fukami2020convolutional,xu2020multi,lee2020model,agostini2020exploration,morimoto2021convolutional}. CAEs are often combined with a time series prediction approach to model the evolving dynamics of the latent space \cite{maulik2021reduced}. A self-attention mechanism is also used within the CAEs to enhance the feature extraction ability of the network \cite{wu2021reduced,fu2021data}. A combination of principal component analysis (PCA) with Gaussian process regression has been investigated in the development of digital twins of reacting flow applications \cite{aversano2019application}. These approaches bypass the Galerkin projection and hence become fully nonintrusive \cite{pawar2019deep,rahman2019nonintrusive,yu2019non,dutta2021pynirom}, building an end-to-end ROM solely from available data without requiring any knowledge of the underlying evolution equations. However, the number of trainable parameters might quickly reach millions even for two-dimensional PDE settings \cite{pawar2020priori}. Moreover, the padding approaches that are available in most CAE architectures might pose additional challenges in representing the desired boundary conditions \textcolor{rev1}{unless specific customization is embedded}. 


In this letter, we introduce a simple and modular nonintrusive Galerkin-free ROM framework for unsteady flows by employing ideas from POD-based and AE-based ROM approaches (see the bottom panel in Figure~\ref{fig:nonPOD}). In our methodology, we first apply the POD procedure to generate a set of orthonormal basis functions. Instead of truncating the number of modes, we consider an almost full-rank POD expansion defined by 99.9\% of the relative information content (RIC) measure, and perform an inner product between snapshots and the POD modes to obtain the set of POD coefficients. We then construct a plain multilayer perceptron AE for finding the embedding of these POD coefficients (which requires substantially fewer trainable parameters compared to those required by the convolutional AEs) to generate a nonlinear mapping between these POD coefficients and a latent space constructed with only a few parameters in the bottleneck layer. Finally, we utilize a recurrent neural network approach for integrating evolution dynamics of the latent space. In this modular way, we utilize the best features of all relevant approaches. Specifically, POD is utilized in ranking the structural content using a linear spatiotemporal decomposition, AE is used in nonlinear dimension reduction, and LSTM is integrated for the time series prediction. Following the established name convention of nonlinear PCA (NLPCA) \cite{kramer1991nonlinear,monahan2000nonlinear,hsieh2001nonlinear,hsieh2007nonlinear}, we call our approach nonlinear POD (NLPOD). 

\textcolor{rev1}{We note that previous studies aimed at leveraging the synergy between POD and AEs. One notable example is their use together with dynamic mode decomposition approaches \cite{otto2019linearly,erichson2019physics,iwata2020neural}. In particular, the combination of POD with AE has been exploited to efficiently reveal Koopman invariant subspaces \cite{pan2020physics}. Furthermore, the idea of using POD and AE to develop a constrained Koopman neural network framework has been proposed by Puligilla and Jayaraman \cite{puligilla2018deep}. Although most ROM efforts aim at providing a predictive model trained with ample amount of data with carefully-designed experiments to be operated in real-time, ROM can also act as a method for data compression to mitigate challenges related to data storage and transfer. For instance,  Carlberg {\it et al.} \cite{carlberg2019recovering} applied PCA to reduce the dimensionality of the vector of local encodings across the entire spatial domain to facilitate data I/O in large-scale simulations.}

\textcolor{rev1}{The combination of POD/PCA and AE has been also utilized for model order reduction, similar to the NLPOD methodology. Casas, Arcucci, and Guo \cite{casas2020urban} retained all the principal components (PCs) of urban air pollutant data and applied a fully connected AE onto these PCs for further reduction. They found that the PC-AE requires substantially fewer trainable parameters than an equivalent fully connected AE on the full-space. This approach was adopted by Phillips {\it et al.} \cite{phillips2021autoencoder} with an additional step of linearizing the decoder mapping from the latent variables to the scalar fluxes to reduce the computational costs of solving eigenvalue problems using the power iteration method. This combination was also employed by Kosut, Ho, and Rabitz \cite{kosut2021quantum} in the quantum physics community as a better alternative than using a plain-vanilla AE approach.}

\textcolor{rev1}{In the present study, we apply AE onto the coefficients of a RIC-guided POD expansion of parameterized convection-dominated flows. The truncation of basis function beyond a prespecified RIC value helps to filter out redundant and noisy signals before passing them to the AE. It also provides an informed trade-off between accuracy and storage requirements. Furthermore, we discuss in detail the benefits of using POD as an upper layer for the interface with the physical space, compared to CAEs. This includes computational aspects (e.g., training and deployment costs) and physical insights (e.g., respecting boundary conditions, symmetries, and conservation laws). Furthermore, we explore the applicability of NLPOD for time-dependent parametric systems, where we augment the bottleneck layer of the autoencoder with the operating parameters to enhance the latent space discovery. We highlight that NLPOD provides a robust end-to-end ROM data compression framework built using significantly fewer trainable parameters than CAEs. Moreover, NLPOD can easily handle data on unstructured and non-uniform grids. Our approach could be well suited for the digital twin applications, where fast transfer learning procedures are often desired when new training data streams are incorporated \cite{rasheed2020digital}.}

\section{GPOD modeling}
In standard Galerkin-based ROMs, a handful of orthonormal basis functions are predetermined with the assumption that the solution approximately lives in the subspace spanned by these bases. POD provides a systematic way for building such basis functions from collected data sets (often called snapshots). Arguably, the simplest way to perform POD involves the singular value decomposition (SVD). For instance, assuming that the state variable is denoted by $u(\mathbf{x},t) \in \mathbb{R}^{N}$ (where $N$ is the spatial resolution), a data matrix $\mathbf{A} \in \mathbb{R}^{N\times m}$ can be formed by stacking $m$ temporal snapshots as $\mathbf{A} = [u(\mathbf{x},t_1), u(\mathbf{x},t_2), \dots, u(\mathbf{x},t_m)]$. Then, an SVD can be applied either directly onto $\mathbf{A}$ (or onto its mean-subtracted/shifted version) as follows:
\begin{equation}
    \mathbf{A} = \mathbf{U} \boldsymbol{\Sigma} \mathbf{V}^*,
\end{equation}
where $\mathbf{U}$ and $\mathbf{V}$ are orthogonal matrices representing the left and right singular vectors of $\mathbf{A}$, respectively, while $\boldsymbol{\Sigma}$ is a diagonal matrix containing the corresponding singular values, $\sigma_k$, in descending order. The columns of $\mathbf{U}$ define the sought POD modes, $\mathbf{U} = [\phi_1, \phi_2, \dots]$. Moreover, the inherent sorting of the singular values provides a valuable feature of POD since the most important modes are placed first and the RIC values of the leading $r$ modes can be defined as
\begin{equation}
    \text{RIC} (\%) = \dfrac{\sum_{k=1}^{r} \sigma_k^2}{\sum_{k=1}^{m} \sigma_k^2} \times 100,
\end{equation}
where $\sigma_k$ is the $k^{th}$ singular value. Then, a rank-$r$ approximation of the state variable can be written as a linear superposition of the contributions of these first $r$ modes as follows:
\begin{equation}
    u(\mathbf{x},t) = \sum_{k=1}^{r} a_k(t) \phi_k(\mathbf{x}) \label{eq:urom}.
\end{equation}
In order to compute (evolve) the time-dependent coefficients $a_k(t)$, Galerkin-based methods perform a projection of the governing equations onto the corresponding POD modes to yield a system of $r$ ordinary differential equations for the vector of coefficients $a_k$ as follows:
\begin{equation}
    \dfrac{d \boldsymbol a}{d t} = f(\boldsymbol a).
\end{equation}
Due to the quadratic nonlinearity in most fluid flow systems, the \emph{online} computational cost of the Galerkin POD (GPOD) approach scales cubically with the number of retained modes. Thus, a modal truncation is performed to achieve a computational gain from GPOD. Nonetheless, most convection-dominated flows cannot be simply described using Eq.~\ref{eq:urom} if $r$ is small, which leaves us working in under-resolved regimes, where there are both a representation error and a closure error \cite{ahmed2020long}.

\section{CAE modeling}
In order to mitigate the deficiency of POD in providing a \emph{compact} set of basis function, CAE frameworks have shown substantial success in providing a compressed latent space that defines a nonlinear manifold on which the system's dynamics evolves more accurately than on the linear manifold defined by a similar POD compression. The CAE starts with an encoding process that involves applying a series of convolutions and nonlinear mappings onto the input snapshot data to shrink the dimensionality down to a bottleneck layer representing the low rank or latent space embedding. An inverse mapping from the latent space variables to the physical space is performed by another set of deconvolutions and nonlinear mappings, defining the decoding part. \textcolor{rev1}{For example, denoting the encoder function as $\eta$ and the decoder as $\zeta$, we can represent the manifold learning through autoencoder as follow,
\begin{align}
     \eta, \zeta  = & \ \underset{\eta, \zeta}{\arg\max} \| \mathbf{A} - (\eta \circ \zeta)(\mathbf{A}) \|, \\
     \eta: & \ u(\mathbf{x},t) \in \mathbb{R}^N \to \boldsymbol z \in \mathbb{R}^r, \\    
     \zeta: & \ \boldsymbol z \in \mathbb{R}^r \to u(\mathbf{x},t) \in \mathbb{R}^N, \label{eq:ae}
\end{align}
where $\boldsymbol z$ represents the low order code at the bottleneck. Note that $\eta$ and $\zeta$ are parameterized by the neural network weights and biases, which are learned in the training process.}

For the temporal dynamics, a surrogate model emulator is constructed to evolve the latent variables onto the manifold revealed by the CAE. 
In the present study, we utilize the capabilities of the long short-term memory (LSTM) networks in sequential data prediction to propagate the latent variables in time. Despite the success and popularity of CAEs in the past few years, there still exist potential limitations to their applications in fluid flows. Below, we highlight a few of them.

\begin{itemize}
    \item The number of trainable parameters quickly explodes with the dimensionality of input data and complexity of the CNN architecture. As a result, proper training requires a prohibitively large number of training samples and a long training time.
    \item The design of CNNs often involves padding operations, and physically-consistent boundary conditions are not guaranteed unless custom padding is adopted (e.g., for periodic boundary conditions). Thus, it is more challenging to impose general boundary conditions in CAEs.
    \item It is rarely possible to physically and/or mathematically interpret the CAE latent space. Therefore, the use of CAE to extract and analyze the underlying coherent structures is limited.
    \item Plain CAEs do not guarantee conservation properties (e.g., continuity) unless additional constraints are explicitly imposed \cite{mohan2020embedding}.
    \item There are no clear estimates for error bounds given a latent space dimension, and the output totally depends on the architecture hyperparameters and the training process. Thus, it is not possible to evaluate how much accuracy is gained/lost by varying the dimensionality of the latent space. Similarly, there is no information regarding the relevant importance of different latent variables.
    \item \textcolor{rev1}{CNNs have been intrinsically designed to process data that are uniformly sampled in the spatial domain. However, in computational disciplines, we often encounter data that are defined on non-uniform grids. Plain-vanilla CAEs cannot be applied in these cases unless a mapping of the original data onto a uniform grid is employed \cite{gao2021phygeonet}.}
\end{itemize}

\section{NLPOD modeling}
In the proposed NLPOD framework, we aim at utilizing the capabilities of both the GPOD and CAE modeling approaches while mitigating their potential limitations. In particular, we are interested in promoting the following key benefits of the POD methodology:
\begin{itemize}
    \item There is a rich history of POD developments for large scale problems, which makes the compression step more computationally efficient than training CAE, e.g., by using randomization and streaming algorithms.
    \item By construction, the POD basis functions respect the underlying boundary conditions prevailing in the input data sets. Therefore, the reconstructed fields from POD are supported by physically consistent boundary conditions.
    \item The flow fields reconstructed using POD satisfy the conservation laws, which improves the physical soundness of the resulting ROM.
    \item There exists a clear mathematical and physical definition of the resulting POD modes, which renders POD as a useful tool for further analysis and interpretation of the dominant patterns \cite{rathinam2003new}.
    \item Each mode is associated with a metric of its relative importance and contribution to the overall information/variance in the given data sets. Therefore, error bounds have been proved for a given level of data compression.
    \item \textcolor{rev1}{POD is not restricted to specific representation of the flow field variables. For example, POD can be easily extended to data defined on non-uniform or unstructured grids by employing different types of inner products \cite{xie2020lagrangian} and numerical integration schemes.}
\end{itemize}
However, in order to accurately describe convection-dominated flows, a large number of POD modes is often required, which increases the computational cost of solving the ROM if a Galerkin projection is used. We mitigate this limitation by inserting a second layer of compression that uses a standard feed-forward neural network AE for the POD coefficients. In particular, we first employ the standard POD algorithm to efficiently decrease the dimensionality of the input field $u(\mathbf{x},t)$ to \emph{near} full-rank approximation as follows:
\begin{equation}
    u(\mathbf{x},t) = \sum_{k=1}^{n} a_k(t) \phi_k(\mathbf{x}) \label{eq:urom2},
\end{equation}
where $n$ is the number of modes that capture the required amount of information (e.g., $\text{RIC} = 99.9\%$). Then, we train the AE to learn the underlying nonlinear correlations between the coefficients $\{a_k(t)\}_{k=1}^{n}$ to yield a further encoded latent space $\boldsymbol z \in \mathbb{R}^r$, where $r \ll n$. Next, we exploit a simple LSTM architecture to evolve $\boldsymbol z$ forward. Finally, we utilize the decoder part of the trained AE to recover the full-rank POD expansion coefficients. With this, we provide an end-to-end Galerkin-free ROM that enables us to efficiently preserve the aforementioned merits of POD. Moreover, the AE component of the NLPOD is trained onto the POD coefficients, so its architecture becomes independent of the dimensionality of the full order data. Therefore, training NLPOD is significantly more efficient than training CAE directly onto the input field data.

\section{Results and discussion}
We showcase the performance of the NLPOD framework using the Marsigli flow problem, where a fluid is placed into two partitions with different temperatures. The separating barrier is removed instantaneously to allow the fluids to slide over each other in a convection-dominated and buoyancy-driven manner. More details on the problem setting can be found in Refs.~\cite{san2015principal,ahmed2021multifidelity}. In order to decrease the computational cost of training the CAE, in the present study, we consider a uniform Cartesian grid of $512 \times 64$. \textcolor{rev1}{We collect spatio-temporal data corresponding to different values of Reynolds number ($\text{Re})$ to provide a parameterized time-dependent setup. In particular, we store 200 snapshots at each value of $\text{Re} \in \{700,900,1100,1300\}$ and use them for the offline training phase (i.e., $m = 800$). We then test the performance of GPOD, CAE, and NLPOD at $\text{Re} = 1000$, which is not included in the training data set. In order to improve the accuracy of AEs, we augment the bottleneck layer with the Reynolds number as an additional input feature. In other words, the decoder function for the CAE (defined in Eq.~\ref{eq:ae}) becomes $\zeta: \boldsymbol z \in \mathbb{R}^r \cup \text{Re} \in \mathbb{R} \to u(\mathbf{x},t) \in \mathbb{R}^N$. For the NLPOD, it is rather defined as $\zeta: \boldsymbol z \in \mathbb{R}^r \cup \text{Re} \in \mathbb{R} \to \boldsymbol a \in \mathbb{R}^n$. We train the AEs and LSTMs separately for the sake of simplicity and to facilitate the CAE and NLPOD combinations with different time series prediction tools.} 

\begin{figure}[ht]
\centering
\includegraphics[width=0.6\linewidth]{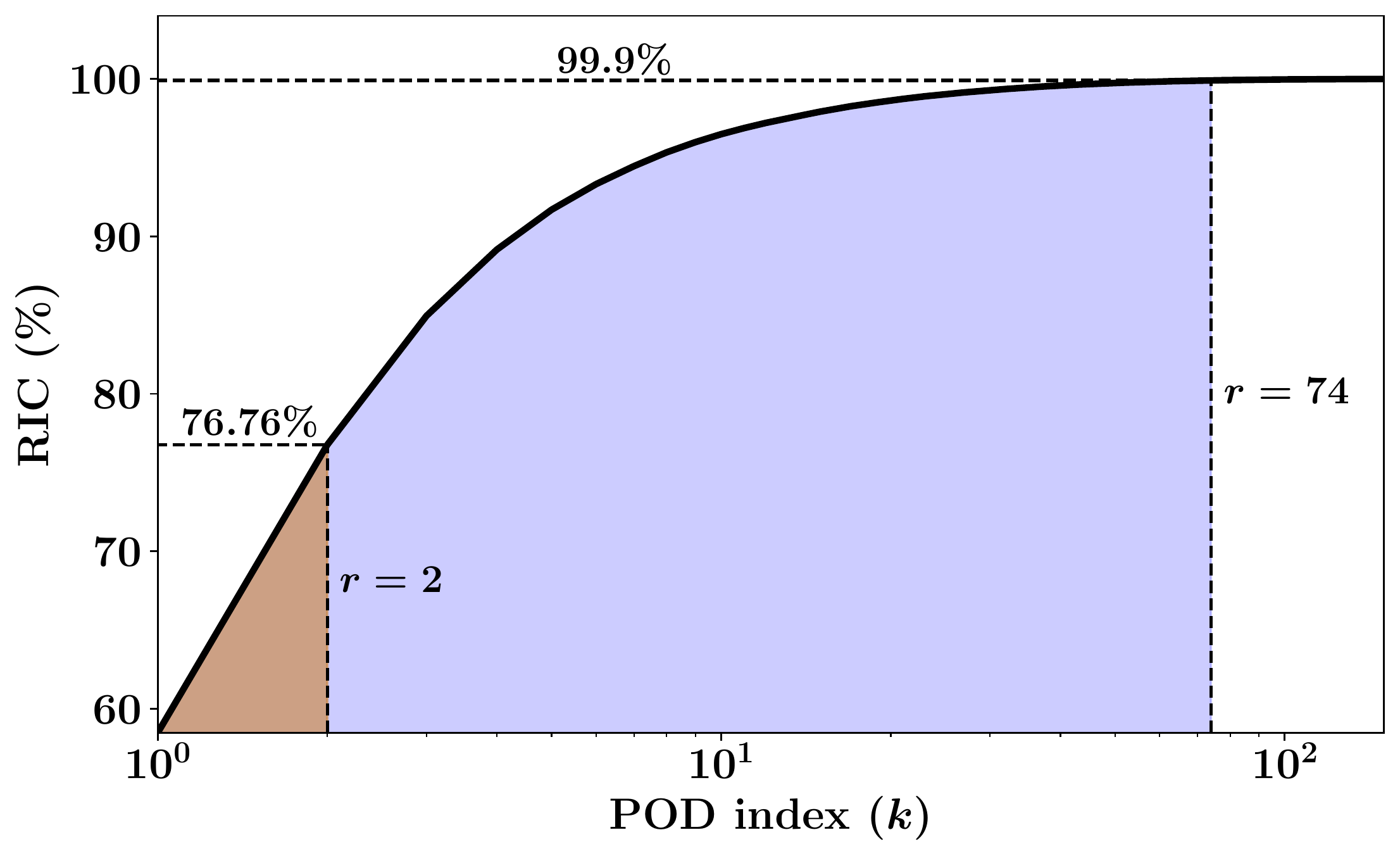}
\caption{The relative information content for different numbers of retained modes. While the first $2$ POD modes have only a $76.76\%$ RIC, the leading $74$ POD modes have a $99.9\%$ RIC of the spatiotemporal temperature field data. Therefore, the proposed autoencoder is built to learn a latent space from $74$ POD coefficients.}
\label{fig:ric}
\end{figure}

\begin{figure*}[ht]
\centering
\includegraphics[width=0.99\linewidth]{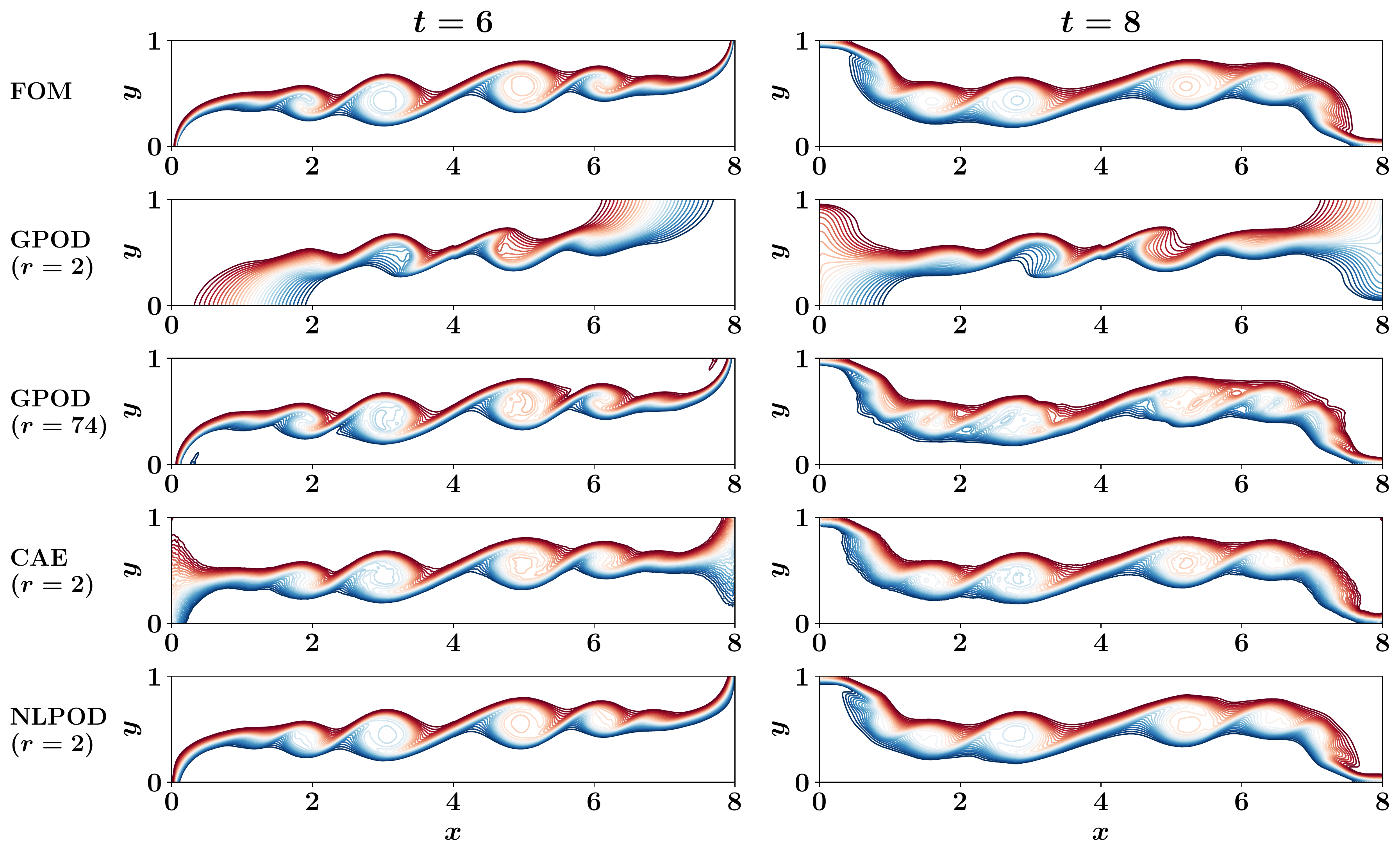}
\caption{Predictions of temperature fields using FOM, compared to GPOD, CAE, and NLPOD modeling approaches. Training NLPOD is significantly faster than training CAE while the computational efficiency (running time) of NLPOD becomes in the same order with GROM ($r=2$), and it is significantly faster than GROM ($r=74$). Note that CAE can yield physically-inconsistent results and fails to capture the accurate boundary conditions. \textcolor{rev1}{Outputs of CAE and NLPOD represent the ensemble-averaged predictions from 10 different neural network weight initializations.}}
\label{fig:contours}
\end{figure*}

We aim at approximating the temperature fields and assume that our target low order compression is $r=2$. For the given problem, Figure~\ref{fig:ric} shows that $2$ POD modes capture only $76.76\%$ of the information, and at least $74$ POD modes are required to attain a RIC value of $99.9\%$ (i.e., rank-$n$ approximation with $n = 74$). In the top panels of Figure~\ref{fig:contours}, we plot the temperature fields at $t=6$ and $t=8$ from the FOM solver and the GPOD results for $r=2$ and $r=74$. Although the GPOD provides good accuracy with $r=74$, the online computational cost is relatively high (around $10$ minutes, compared to a fraction of a second for GPOD for $r=2$). Therefore, we train AE onto the $74$ POD expansion coefficients $\boldsymbol a(t) \in \mathbb{R}^{74}$ to learn a zipped latent space for $\boldsymbol z(t) \in \mathbb{R}^{2}$. Similarly, we train a CAE to reduce the dimensionality of the temperature field from $u(\mathbf{x},t) \in \mathbb{R}^{512 \times 64}$ to a latent space for $\boldsymbol z(t) \in \mathbb{R}^{2}$. 

The reconstructed temperature fields from NLPOD and CAE frameworks are shown in Figure~\ref{fig:contours}, where we can see that the NLPOD predictions are more accurate than the CAE predictions. \textcolor{rev1}{In order to provide an estimate of the model uncertainty in the output results, we utilize an ensemble-based framework, where we train different networks with different weight initializations (i.e., by changing the seed number). We find that the latent space revealed by CAE varies significantly with the initial weights specification. Therefore, we perform the training with 30 different initializations, and select 10 cases, corresponding to the highest performance on the validation data sets, to build our ensemble.} We can observe the non-smoothness of the contour lines at different places in the CAE results, which indicates a non-physical behavior and/or insufficient latent space. Moreover, there is an evident problem in reconstructing the boundary conditions with CAE. \textcolor{rev1}{Although this boundary condition issue can be reduced with proper padding methods, it requires additional case-specific customization of the neural network architecture.} On the other hand, these complications are naturally mitigated in the NLPOD results since the restriction step and the prolongation process back to the FOM space are performed under the physically accurate POD umbrella. We also emphasize that in our experiments the training time of the CAE is significantly larger than the training time of the NLPOD (even using GPU capability), while the online deployment time for both cases is around half a minute. Thus, as shown in Table~\ref{table:MSE}, NLPOD is both more accurate and more efficient than the CAE. \textcolor{rev1}{Moreover, standard deviation data obtained from an ensemble of 10 different runs (i.e., runs trained using different weight initializations) clearly show that the uncertainty of the reconstruction is significantly lower in the NLPOD approach than in the CAE approach. Quantitatively, the standard deviation of the CAE reconstructions is about 55\% of their mean value, while the standard deviation of the NLPOD reconstructions is only 4\% of their mean value.}

\begin{table*}[htbp!]
\caption{Comparison of performance metrics for the NLPOD and CAE methodologies. Training times are reported only for autoencoder part since LSTM training and deployment times are similar in both cases. We perform numerical experiments using two different versions of tensorflow without$^*$ and with$^{**}$ GPU support. \textcolor{rev1}{The data in the table correspond to the mean value $\pm$ the standard deviation, obtained from an ensemble of 10 different runs.} }
\centering
\begin{tabular}{p{0.35\textwidth} p{0.2\textwidth} p{0.2\textwidth}  p{0.18\textwidth} }  
\hline
Model & CAE & NLPOD & Gain [CAE/NLPOD] \\
\hline
Number of trainable parameters & 2,194,499 & 40,516 & 54.16 \\
CPU training time [s]$^*$ & \textcolor{rev1}{$1239.05 \pm 32.30$} &  \textcolor{rev1}{$14.16\pm0.44$} & \textcolor{rev1}{87.48} \\ 
GPU training time [s]$^{**}$ & \textcolor{rev1}{$123.30\pm 2.09$} & \textcolor{rev1}{$56.2 \pm 0.80$} & \textcolor{rev1}{$2.19$}\\
MSE of reconstruction (without LSTM) & \textcolor{rev1}{$(7.18 \pm 3.93) \times 10^{-4}$} &  \textcolor{rev1}{$(1.25\pm 0.05) \times 10^{-4}$} &  \textcolor{rev1}{5.75}\\
MSE of prediction (with LSTM) &  \textcolor{rev1}{$(1.53 \pm 0.75) \times 10^{-3}$} &  \textcolor{rev1}{$(3.61\pm 0.74) \times 10^{-4}$} &  \textcolor{rev1}{4.24} \\
\hline \smallskip  \\
\end{tabular}
\label{table:MSE}
\end{table*}

\section{Conclusions}
In conclusion, CAE technology has been shown to be a crucial dimensionality reduction approach for nonintrusive modeling and prediction of fluid flows. However, the CAE training time could be significant for high-dimensional systems often encountered in fluid dynamics applications. On the other hand, the intrusive projection-based approach has been the workhorse principle for reduced order modeling of fluid flows. The governing equations of fluid flows possess a quadratic nonlinearity. Therefore, POD-based methods, often combined with the Galerkin projection, yield a coupled dynamical system with computational complexity of $O(r^3)$, where $r$ is the number of retained modes. To become computationally tractable, these projection-based methods truncate the reduced order representation ($r<n$), where $n$ is the full rank of snapshot data matrix. Unfortunately, this truncation leads to a large projection error, especially for convection-dominated flows. In this work, we introduce a robust nonintrusive method that combines POD and multilayer perceptron autoencoders to generate a  projection-error-free reduced order representation (with significantly reduced training time), and integrate this latent space with LSTM recurrent neural networks for its dynamics. Our results for modeling a lock-exchange density current problem show a substantial performance improvement over both nonintrusive convolutional AEs and intrusive Galerkin ROMs. 

This is the first step toward building robust end-to-end Galerkin-free nonintrusive models for convection-dominated flows. Our future efforts will aim at extending this autoencoder-based framework to more complex higher-dimensional multiphysics problems in fluid dynamics. We highlight that advanced hyperparameter selection studies may be conducted for performance boost. Furthermore, instead of a deterministic autoencoder, one can utilize a variational autoencoder to perform probability distribution modeling of the latent space. \textcolor{rev1}{It is also worth noting that the NLPOD approach is not restricted to a specific methodology of time series prediction. Although we introduce our results using LSTMs, NLPOD is equally applicable with other techniques \cite{bai2021non}, including sparse regression \cite{fukami2021sparse}, Gaussian process regression\cite{maulik2021latent}, Seq2seq algorithms \cite{reddy2019reduced}, temporal fusion transformers \cite{lim2021temporal}, and auto-regression methods. One can also replace LSTM with neural ODEs \cite{NEURIPS2018_69386f6b} to accommodate varying time stepping discretization.} Lastly, the proposed NLPOD ROM framework can be seamlessly integrated with the physics-guided machine learning \cite{pawar2021model} to reduce the uncertainty of the nonintrusive model.



\section*{Acknowledgements}
This material is based upon work supported by the U.S. Department of Energy, Office of Science, Office of Advanced Scientific Computing Research under Award Number DE-SC0019290. O.S. gratefully acknowledges the Early Career Research Program (ECRP) support of the U.S. Department of Energy. O.S. also gratefully acknowledges the financial support of the National Science Foundation under Award Number DMS-2012255. T.I. acknowledges support through National Science Foundation Grant Number DMS-2012253.

\section*{Data availability}
The data that supports the findings of this study are available within the article. Implementation details
and Python scripts can be accessed from the Github repository\cite{git}.

\bibliographystyle{unsrt} 

\bibliography{manuscript}

\end{document}